\documentclass[10pt]{iopart}
\begin{document}
\title[Stresses in lipid membranes]{Stresses in lipid membranes}
\author{R Capovilla\dag and
J Guven\ddag}
\address{\dag\
Departamento de F\'{\i}sica,
Centro de Investigaci\'on y de Estudios Avanzados del IPN,
 Apdo. Postal 14-740, 07000 M\'exico DF, MEXICO}
\address{ \ddag\ Instituto de Ciencias Nucleares,
Universidad Nacional Aut\'onoma de M\'exico,
 Apdo. Postal 70-543, 04510 M\'exico DF, MEXICO}

\begin{abstract}
The stresses in a closed lipid membrane 
described by the Helfrich hamiltonian, quadratic in the extrinsic curvature,  
are identified using Noether's theorem. 
Three equations describe the conservation of the stress tensor: 
the normal  projection is identified as 
the shape equation describing equilibrium configurations; 
the tangential projections are consistency conditions on the 
stresses which capture the fluid character of such membranes.
The corresponding torque tensor is also identified. 
The use of the stress tensor as a basis for perturbation theory is discussed.
The conservation laws are cast in terms of the forces and torques on closed
curves.  
As an application, the first integral of the shape equation 
for axially symmetric configurations is derived by examining the 
forces which are balanced along circles of constant latitude.
\end{abstract}

\pacs{87.16.Dg, 46.70.Hg}


%\maketitle

\section{Introduction}

The Helfrich hamiltonian, quadratic in the extrinsic curvature, provides a 
remarkably
robust description of the mechanical properties of lipid
membranes (see, for example, \cite{Can:70,Hel:73,Nel.Pir.Wei:89,Saf:94,Pel:94,Boa:02}, and the
comprehensive review \cite{Sei:97}).  This hamiltonian associates an
energy penalty with bending. It is invariant with respect to deformations
along the membrane itself.  Equilibrium configurations will therefore
satisfy a single `shape' equation corresponding to the 
extrema of the hamiltonian with respect to a deformation
directed along the normal\cite{Hel.OuY:87,Hel.OuY:89}.
This equation is purely geometrical. 
Underpinning any non-trivial geometry, however, there will always
be stresses within the membrane.
In this paper, we will focus on the distribution of these stresses, as well
as the corresponding torques, and how the two are reflected in the geometry.
Apart from its conceptual value, it might appear that
the formulation of the problem this way
represents a step in the wrong direction; after all,
the shape equation describes the physics completely,
and to obtain it, there is no need to know what the stresses are, 
much less the torques.  It is surprising then, that by stepping back a little from the 
geometry this way, a potentially
powerful as well as economical approach to the physics is also provided.
We should remark that the idea itself in this context is not new, 
going back to the early work by Evans \cite{Eva:74}. More recently, it has also
been exploited for axisymmetric geometries in
\cite{GHP:02} to study fluid membrane tethers.

TTo construct the stress tensor we appeal to Noether's theorem:
the euclidean invariance of the Helfrich hamiltonian implies conservation
laws
and corresponding conserved charges. (For an
analogous treatment of elastic curves in space see \cite{Cap.Chr.Guv:02a}.) 
In particular, translational invariance implies the proportionality of the 
divergence of the membrane stress tensor 
to any applied external forces; if the membrane is closed and the 
enclosed volume is fixed, this source  will be the hydrostatic pressure
enforcing the constraint. If the membrane is acted on by a localized 
external force, as is the case of micro-manipulation techniques 
\cite{Hel.Har.Bou:00,Hoc:00},
this will appear as an additional distributional source on the membrane.

Implicit in Noether's theorem is the recipe for the construction 
of the relevant conserved quantities. This is important 
because the corresponding Newtonian construction, while simple in principle, 
is {\it not} in practice. (See, however,  \cite{GHP:02},
where the stresses and the torques are derived for axisymmetric 
configurations.)

The relationship between the three equations describing the 
conservation of the stress tensor and the (single) shape equation 
is very simple:
the projection onto the normal of the former is the shape equation. 
The projection process necessarily dismantles the 
divergence form of these equations:
the divergence of the normal stress
picks up a source proportional to the local tangential stress, the 
`extrinsic curvature cubed'  nonlinearity in the shape equation.  
The fluid character of a lipid membrane is captured mathematically in
the reparametrization invariance of the hamiltonian describing it.
The two tangential projections of the conservation law 
(or its `Bianchi identities' in the language of gauge theory)  
reflect this invariance. As such, these latter two
equations are satisfied even when the shape equation is not. 
This structure itself is model independent. 
There is thus little or no cost involved in 
considering a general hamiltonian depending on the extrinsic curvature, in
particular, models higher order in curvature.

It is also possible to cast the conservation of the stress tensor 
as a more intuitive global statement 
relating the total force on any closed loop lying on the surface
to the action of external forces on the area enclosed by the loop.  
This global statement of conservation is particularly useful when the 
membrane possesses some level of spatial symmetry. 
The first integral of the shape equation 
for axially symmetric configurations of both spherical and 
toroidal topology is obtained as an immediate consequence when the loop
coincides with a circle of constant latitude. This approach to the problem 
contrasts favorably with `hamiltonian' approaches
\cite{Zhe.Liu:93,Boz.Sve.Zek:97}.
We also identify a global constraint on the
tangential stress over the closed membrane, which reproduces the known scaling
identity for the Helfrich hamiltonian \cite{Sei:97}.

The paper is organized as follows. In Section 2 we describe
how the hamiltonian responds to a deformation of the membrane surface.
In Section 3 it is shown how Noether's theorem may be 
applied to identify the conservation of the stress tensor as the
conservation law associated with translation invariance of 
the two-dimensional surface. 
In this section, we also show how the conservation of the stress tensor can be 
projected to cast the shape equation in terms of derivatives of the
stress tensor components. The special case of  
a soap bubble is used in Section 4 to illustrate our approach. 
In Section 5, we extend our considerations to the Helfrich hamiltonian 
and more generally to hamiltonians that depend on  
any power of the extrinsic curvature.
We derive explicit expressions for the stress tensor. In Section 6, we
consider rotational 
invariance and the definition of torque. We introduce an intrinsic torque which, 
added to the couple due to the stress tensor, is conserved.
In Section 7,  the global form of the 
conservation laws is examined; a non-trivial global relationship
which involves the stress tensor itself is identified. In Section 8 we
comment briefly on how the stress tensor might be exploited to develop 
perturbation theory. In Section 9, we specialize to axially symmetric
configurations.
We show how to obtain a relationship between the components 
of the stress tensor that encodes the shape equation. 
We conclude with some brief remarks in Section 10. 

\section{Response of the hamiltonian to surface deformations}

Let us consider an embedded surface $\Sigma$
in three-dimensional euclidean space.
For simplicity, we suppose that this surface is closed.
The surface is described locally by three functions 
${\bf x} = {\bf X}(\xi^a)$,
where ${\bf x}=(x^1,x^2,x^3)$ are for now cartesian coordinates
on $R^3$,   $\xi^a, a=1,2$ may be any local coordinates
on the surface. We introduce two tangent vectors  $ {\bf e}_a$ which are defined by 
${\bf e}{}_a= \partial{\bf X}/\partial \xi^a$.
The surface geometry is described completely by the induced 
metric $\gamma_{ab}$, and the extrinsic curvature $K_{ab}$. 
The former is given by
\begin{equation}
 \gamma_{ab} = {\bf e}{}_a
\cdot {\bf e}{}_b \,.
\end{equation}
We denote by $\nabla_a$
the covariant derivative on the surface which is compatible with $\gamma_{ab}$.
The  extrinsic curvature $K_{ab}$ is given by 
\begin{equation}
K_{ab} = - {\bf n}\cdot \partial_a {\bf e}{}_b\,,
\label{eq:ec}
\end{equation}
where ${\bf n}$
is the  unit vector normal to the surface.
We will indicate the trace of $K_{ab}$ by $K = \gamma^{ab} K_{ab}$.
The intrinsic and extrinsic geometries are related by the Gauss-Codazzi-Mainardi equations,
\begin{eqnarray}
K^{a b}K_{a b} - K^2 -  {\cal R} &=& 0\,, \label{eq:gc}\\
\nabla_a K_{bc} - \nabla_b K_{ac} &=& 0\,, \label{eq:cm}
\end{eqnarray}
where ${\cal R}$ denotes the surface scalar curvature.

Though our principal interest in is the case of a lipid membrane,
we will consider any local hamiltonian, depending on the
functions ${\bf X}$, which is 
invariant under surface reparametrization; no energy penalty 
is associated with shearing deformations  of the membrane. 
The surface hamiltonian is given by 
\begin{equation}
H_\Sigma [{\bf X} ] = \int_\Sigma dA\; \; h \;(\gamma_{ab}, K_{ab},
\nabla_a K_{bc}, \cdots)
 \,.
\label{eq:h}
\end{equation}
where the scalar $h$  is constructed locally
from the geometry of the surface, and the infinitesimal
area element is given by
 \begin{equation}
dA = d^2\xi \; \sqrt{\gamma}\,,
\end{equation}
with $\gamma = \mbox{det} \,\gamma_{ab}$.
In particular, the Helfrich hamiltonian
is proportional to 
\begin{equation}
H_{(2)} =  \int_\Sigma dA \; K^2 \,.
\label{eq:helfrich0}
\end{equation} 
If there is a spontaneous curvature $C_0$,
$K$ is replaced by $K-C_0$ in Eq.(\ref{eq:helfrich0}).
If the membrane is closed there may be a contribution to the total
hamiltonian
proportional to 
the enclosed volume $V$ due to a constant pressure excess $P$ there.
In addition, there may be global constraints imposed on the 
membrane geometry. For example, one or more of the area, the integrated mean curvature,
\begin{equation}
M = \int_\Sigma dA\; K \,,
\end{equation}
or the enclosed volume $V$ may be fixed.
In Helfrich's original model, both $A$ and $V$ are fixed. 
In Svetina and \u{Z}e\u{k}'s bilayer couple model, in addition to these
$M$ gets fixed
\cite{Sve.Zek:89}. 
These constraints are imposed by lagrange multipliers.
 We note that in the refinement of these models, known as the 
area difference model, the hamiltonian 
is not expressible in the simple form (\ref{eq:h}):  a 
non-local term proportional to $(M - M_0)^2$ is
added to the hamiltonian, where $M_0$ is some constant
\cite{Sei:92,Wie:92,Boz:92,ADM:93}. In this model,
deviations in the integrated mean curvature from $M_0$ 
are penalized energetically. We will not consider this model further.

To proceed, we need to know how this hamiltonian responds to an 
infinitesimal deformation of the 
embedding functions for the surface,
${\bf X}(\xi) \to  {\bf X}(\xi) + \delta {\bf X}(\xi)$.
We decompose $\delta {\bf X}$ into its parts 
tangential and normal to the surface, 
\begin{equation}
\delta {\bf X}= \Phi^a {\bf e}{}_a + \Phi {\bf n}\,.
\label{eq:vd}
\end{equation}
As an intermediate step, 
we evaluate how the quantities $\gamma_{ab}$ and $K_{ab}$ respond to
this deformation. The tricky bit is to do this in a way 
which does not depend on the particular reparametrization 
of the surface we choose. For the induced metric we have
\begin{eqnarray}
\delta \gamma_{ab} &=&  {\bf e}_{a} \cdot \nabla_{b} \; \delta {\bf X}
 + {\bf e}_{b} \cdot \nabla_{a} \; \delta {\bf X} \nonumber \\
&=& 2  \; K_{ab}\; \Phi + \nabla_a \Phi_b + \nabla_b \Phi_a \,.
\label{eq:defmet}
\end{eqnarray} 
To obtain the second line we have used the decomposition (\ref{eq:vd}),
together with the Gauss-Weingarten equations,
\begin{eqnarray} 
\nabla_a {\bf e}_b &=& - K_{ab} {\bf n} \,, \label{eq:gw1}\\
\nabla_a {\bf n} &=& K_{ab} {\bf e}^b\,. \label{eq:gw2}
\end{eqnarray} 
Similarly, for the extrinsic curvature defined by 
(\ref{eq:ec}), we find 
\begin{eqnarray}
\fl
\delta K_{ab} &=& - \left( \delta {\bf n} \right) \cdot
\nabla_a \nabla_b \; {\bf X}  - {\bf n} \cdot \nabla_a 
\nabla_b \;\delta {\bf X}\nonumber \\ \fl
&=&  - \nabla_a \nabla_b \Phi + K_{ac} K^c{}_b \Phi
+ \Phi^c \nabla_c K_{ab} 
+ K_{ac} \nabla_b \Phi^c +
K_{bc} \nabla_a \Phi^c\,. \label{eq:defcur}
\end{eqnarray}
Here the first term in the first line vanishes, because of the unit 
vector fact:
$ \left( \delta {\bf n} \right) \cdot
\nabla_a \nabla_b  {\bf X} = - K_{ab} \left( \delta {\bf n} \right) \cdot
{\bf n} = 0$. To obtain the third term in the second line we have used
the Codazzi-Mainardi equation (\ref{eq:cm}).
As expected, in both Eqs. (\ref{eq:defmet}), (\ref{eq:defcur}), the tangential deformation is the Lie derivative 
along the surface vector $\Phi^a$.

Once we know how the geometry changes under a surface deformation,
we are in a position to vary any hamiltonian of the form (\ref{eq:h}).
However, we can exploit reparametrization invariance to
simplify matters.
The infinitesimal change in the hamiltonian 
can always be decomposed into its 
tangential and normal  parts, 
\begin{equation}
\delta H = \delta_{\parallel} H + \delta_{\perp} H\,.
\end{equation}
Away from boundaries, the tangential deformation 
can be identified with a reparametrization of $\Sigma$ 
since $\delta_{\parallel}H $ is a boundary term. It is simple to show why this is so:
we have that $\delta_ {\parallel} f 
= \Phi^a \partial_a f$ for any scalar function 
$f(\xi)$ defined on $\Sigma$; in addition,  
under a tangential deformation, the induced metric on $\Sigma$
tranforms as a Lie derivative, as given by the tangential
part of (\ref{eq:defmet}). 
Thus we have that
\begin{equation}
\label{eq:tangdef}
\delta_{\parallel}\sqrt{\gamma} = \partial_a( \sqrt{\gamma}\; \Phi^a)\,.
\end{equation}

The identification of a conservation law will require us to 
isolate a divergence in the variation of the hamiltonian.
Thus let us consider the behavior of the hamiltonian $H_{\Sigma_0}$ when it is 
restricted to some connected domain $\Sigma_0$ with a boundary ${\cal C}$.
In the case of a surface with 
spherical topology, we will suppose without loss of generality that $\Sigma_0$ is also simply connected so that ${\cal C}$ is a 
contractable closed curve. For toroidal or higher genus 
surfaces, one might also be interested in regions which are not simply connected with a non-contractable, disconnected boundary. For notational simplicity, we will proceed as though ${\cal C}$ is also connected. For an arbitrary deformation of $\Sigma_0$ we have 
\begin{equation}
\delta H_{\Sigma_0}= 
\int_{\Sigma_0} d^2\xi\;
\left\{ \left( \delta \sqrt{\gamma} \right) 
\; h 
+ \sqrt{\gamma} 
\; \left( \delta h \right)  \right\}\,.
\label{eq:arbi}
\end{equation}
A tangential deformation of the 
surface thus always results in a pure 
divergence,
\begin{equation}
\delta_{\parallel} H_{\Sigma_0} = \int_{\Sigma_0} d^2\xi\; \partial_a \left(\sqrt{\gamma}
\;  h\; \Phi^a \right)
=  \int_{{\cal C}} ds
\; h \; l_a \;
\Phi^a  \,,
\label{eq:tpda}
\end{equation}
where we have used Stokes theorem in the second
equality, $s$ is arclength along the boundary curve ${\cal C}$,
induced by its embedding in $\Sigma_0$, say
$\xi^a = Y^a(s)$, and $l^a$ is the 
unit normal on ${\cal C}$ pointing out of $\Sigma_0$.

A reparametrization of $\Sigma_0$ can only move its boundary. Thus, if there is no boundary, 
$\delta_{||} H_\Sigma =0$.
If we are interested in a stiff membrane, so that  no 
local change in the area is permitted, then $\Phi_a$ itself must respond to $\Phi$ to
maintain the constraint, $\delta \sqrt{\gamma}=0$. Then, as follows from Eq. (\ref{eq:defmet}), 
$\nabla_a \Phi^a + K \Phi =0$. 
Clearly this implies no constraint for 
a euclidean motion. For general deformations, we only require that the
tangential deformation 
results in a divergence. Thus none of our conclusions are modified.
The enclosed volume also clearly does not change under a surface reparametrization.

Whereas the tangential variation of the hamiltonian
is simple, the normal variation is, in general,
 non-trivial.
However, the latter can always be cast in the form
\begin{equation}
\delta_{\perp} H_{\Sigma_0} =\int_{\Sigma_0} dA \;
\left\{ {\cal E}( h ) \; \Phi
+  \; \nabla_a \;  S^a{}[\Phi] \; \right\}  
\,,
\label{eq:perpdef}
\end{equation}
{\it i.e.} as a bulk part plus a pure divergence. 
Here ${\cal E} (h)$ is the Euler-Lagrange derivative of 
$h$ with respect to surface deformations,
projected onto the normal ${\bf n}$ to the
surface;
$S^a$ is a linear differential operator 
on the surface which operates on the normal deformation $\Phi$ as follows:
\begin{equation}
S^a[\Phi ] = S^a_{(0)}\Phi  + S^{ab}_{(1)} \nabla_b \Phi +  \cdots 
\,.
\end{equation} 
To construct $S^a[\Phi]$, integration by parts is used to collect 
in a pure divergence all surface gradients and higher derivatives
of $\Phi$. Such terms will show up in the normal deformation of $K_{ab}$ and
its derivatives. 

Summing the two independent variations we have
\begin{equation}
\delta H_{\Sigma_0} =\int_{\Sigma_0} dA \;
\left[ {\cal E}( h ) \; \Phi
+ 
 \; \nabla_a \;  Q^a \right]  
\,,
\label{eq:defh}
\end{equation}
where the Noether charge is 
\begin{equation}
Q^a = S^a [ \Phi ]  + h \; \Phi^a\,.
\end{equation}
Note that $Q^a$ is not unique; clearly $Q^a \to Q^a + \epsilon^{ab}
\nabla_b f $, for some scalar density $f$, with $\epsilon^{ab}$ the
surface Levi-Civita density, will leave Eq. (\ref{eq:defh}) unchanged.  

The variational principle restricted to normal 
deformations determines the equilibria.
Suppose that the enclosed volume is fixed at some value $V_0$.
The total hamiltonian 
\begin{equation}
H= H_{\Sigma} - P (V- V_0)
\label{eq:HP}
\end{equation}
is then stationary with respect to normal deformations of $\Sigma$
when the Euler-Lagrange equation 
\begin{equation}
{\cal E} (h)=P
\label{eq:EP}
\end{equation}
is satisfied. This is the `shape' equation determining the equilibria of the membrane.
We have used the fact that the interior volume varies as 
\begin{equation}
\delta_{\perp} V = \int_{\Sigma} dA\;
 \; \Phi\,.
\end{equation}

\section{Stresses in the membrane}

To this point, we have considered arbitrary deformations
of the functions ${\bf X}$. We now ask what occurs 
if we subject the membrane to a translation or a rotation. 
Let us then consider an infinitesimal euclidean 
motion, $\delta {\bf X} = {\bf a}  +{\bf b}\times  {\bf X}$,
where ${\bf a}$ is an infinitesimal
constant translation,  and  ${\bf b}$ represents  
an infinitesimal rotation. Let us focus on translations.
Rotations will be considered in Section \ref{rote}.

If we decompose $\delta {\bf X}$ according to Eq.(\ref{eq:vd}) then for
an infinitesimal spatial translation, $
\delta {\bf X}={\bf a}$, we have $\Phi = {\bf a}\cdot {\bf n}$,
and $\Phi_a = {\bf a}\cdot{\bf e}{}_a$.
Substituting into Eq.(\ref{eq:defh}),  
the variation of the hamiltonian associated
with this  translation can be cast 
in the form,
\begin{equation}
\label{eq:gfdat1}
\delta H_{\Sigma_0}= {\bf a}\cdot
\int_{\Sigma_0} dA\,\left[\; {\cal E} (h)   \;  {\bf n}
-  \nabla_a {\bf  f}^{a}
\right]
\,.
\end{equation} 
The surface vector ${\bf f}^{a}$ is given by 
\begin{equation}
{\bf f}^{a} = -
{\bf S}^a [ {\bf n}] -  h\, {\bf e}^{a}
\,,
\label{eq:pam}
\end{equation} 
where ${\bf S}^a [{\bf n} ]$ is defined by 
\begin{equation}
S^a [\Phi ] = {\bf a}\cdot {\bf S}^a [{\bf n}]\,.
\label{eq:sa}
\end{equation} 
The quantity ${\bf f}^a$ describes the non-vanishing 
components of the stress tensor as appropriate for a two-dimensional system:
it is both a spatial vector and a surface vector. The 
ambiguity in the Noether charge $Q^a$ translates into
an ambiguity in the stress ${\bf f}^a \to {\bf f}^a +
\epsilon^{ab} \nabla_b {\bf W}$, for some arbitrary vector density ${\bf W}$.

While tangential deformations do not participate
in the variational derivation of the shape equation,
we see that they do contribute in a simple but essential way to the construction 
of the stress tensor. 

Suppose that $H_{\Sigma_0}$ is 
invariant under translations,
$\delta H_{\Sigma_0} =0$.
Because $\Sigma_0$ is  arbitrary in Eq.(\ref{eq:gfdat1}), the integrand vanishes 
pointwise so that
\begin{equation}
\nabla_a {\bf  f}^{a} =  \; 
{\cal E} (h)   \;  {\bf n}\,.
\label{eq:fE}
\end{equation}
If, in addition, the Euler-Lagrange equation for $H$ as given by Eq.(\ref{eq:HP}) is satisfied,
then ${\cal E} (h) =P$ in  Eq.(\ref{eq:fE}).

Let us examine Eq.(\ref{eq:fE})  a little more closely.
We note that there 
are three conservation laws, whereas there is only one
shape equation. To resolve this discrepancy,
we decompose the space vector ${\bf f}^{a}$  
into its tangential and normal parts,

\begin{equation}
{\bf f}^{a} =   f^{ab}  {\bf e}{}_b + 
f^{a} {\bf n} \,.\label{eq:gep}
\end{equation} 
Note that, in general, the surface tensor 
$f^{a b}$ need not be symmetric in its indices.
The surface covariant divergence of ${\bf f}^{a}$ gives 
\begin{equation}
\nabla_a {\bf f}^{a} = \left( \nabla_a f^{ab} 
+  K^b{}_{a}   f^{a} \right){\bf e}{}_b +
\left({\nabla}_a  f^{a} -  
K_{ab}{} f^{ab}  \right) {\bf n}\,,
\label{eq:derden}
\end{equation}
where we have made use of the Gauss-Weingarten 
equations for the surface $\Sigma$, Eqs. (\ref{eq:gw1}), (\ref{eq:gw2})
The surface projections of Eq.(\ref{eq:fE}) 
are therefore given by 
 \begin{eqnarray}
{\nabla}_a f^{a} -  K_{ab}  f^{ab}  &=& 
{\cal E} (h) \,,
\label{eq:em2} \\
\nabla_a  f^{ab} +  K^{b}{}_{a}   f^{a} &=& 0\,.
\label{eq:em1}
\end{eqnarray} 
The first equation expresses the Euler-Lagrange 
derivative ${\cal E}(h)$ in terms of $f^{ab}$ and $f^a$. Using 
${\cal E}=P$, the shape
equation takes the remarkably simple universal form
\begin{equation}
{\nabla}_a  f^{a} -  K_{ab}  f^{ab} 
= P\,.
\label{eq:eom3}
\end{equation}
Note that only the symmetric part of $f^{ab}$ contributes.

The conservation of the stress tensor along tangents to the membrane 
(\ref{eq:em1}) is independent of ${\cal E}$. They 
provide consistency conditions on the components of the stress tensor, 
telling us how the tangential
stress must  respond to a given normal stress whether or not the 
shape equation holds. One can think of these equations 
as the Bianchi identities associated with surface reparametrizations. 
It is also clear that they hold separately 
for each term contributing to $H$.

Our treatment so far has been entirely general:
the only properties of the hamiltonian 
we have used are its reparametrization and  translational invariance. 
We will now evaluate the Euler-Lagrange derivative 
and the stress tensor for specific models.

\section{Soap bubbles}

In order to illustrate the ideas that have been 
developed in the previous sections, 
we consider first the  case
of a surface  dominated by surface tension, {\it e.g.} a soap bubble.
The hamiltonian 
is simply proportional to the area
\begin{equation}
H = \mu \int_\Sigma dA\,,
\end{equation} 
where the constant $\mu$ is the surface tension. 
This is the simplest hamiltonian one can write down
for a surface;
it depends only on the intrinsic geometry 
of the surface.  
The normal deformation of this hamiltonian is given by
\begin{equation}
\delta_{\perp} H = \mu \int_\Sigma dA
\; K \; \Phi\,, 
\end{equation}
where we have used the familiar expression
relating the Lie derivative along the normals
of the  area element of $\Sigma$ to its 
mean extrinsic curvature  $K = \gamma^{ab} K_{ab}{}$, as follows from
Eq. (\ref{eq:defmet}),
\begin{equation}
\delta_{\perp} dA
 = \; K\; dA\,.
\label{eq:sqk}
\end{equation}
The tangential deformation of this hamiltonian is simply
given by Eq. (\ref{eq:tpda}), with $h =  \mu$.

In this geometrical language, the shape equation 
is given by 
\begin{equation}
\label{eq:K0}
\mu K = P\,.
\end{equation}
The surfaces that extremize the hamiltonian have 
constant mean extrinsic curvature.
This is a second-order hyperbolic 
partial differential equation for the 
embedding functions, ${\bf X}(\xi)$.
To bring the Euler-Lagrange equations into a more conventional form, 
using the Gauss-Weingarten equations (\ref{eq:gw1}), 
we have that
$K = - {\bf n}\cdot \Delta {\bf X} $,
where $\Delta $ is the surface laplacian.
The tangential projections of $\Delta {\bf X}$ vanish identically.
We can now peel this expression 
and its tangential counterpart to recover, 
\begin{equation}
\mu \Delta {\bf X} = - P {\bf n}\,.
\end{equation}

In this model there is no surface term arising from 
the normal variation, so that 
the quantity introduced in Eq. (\ref{eq:perpdef})  vanishes
identically, $S^{a} [\Phi ]=0$, and only the tangential 
variation contributes to the Noether charge.
This feature is unique to this hamiltonian.
The invariance of the area hamiltonian under euclidean motions 
gives the stress 
\begin{equation}
{\bf f}^{a} = - \mu \,{\bf e}^{a}\,,
\label{eq:lddng}
\end{equation}
which is not only  tangential but also 
isotropic, $f^{ab} = - \mu  \gamma^{ab}$.

\section{The Helfrich hamiltonian }

Let us now consider the problem of extremizing 
the bending energy of a surface given by
Eq.(\ref{eq:helfrich0}) 
subject to constraints on $V$, $A$ (and possibly $M$).
To implement the constraints on the area
and the integrated mean curvature, we 
construct the constrained hamiltonian
\begin{equation}
H= \alpha H_{(2)} + \beta (M- M_0) + \mu (A-A_0)\,.
\label{eq:helfrich}
\end{equation}
This is a sum of terms of the form
\begin{equation}
H_{(n)} = \int  dA\; K^n \,.
\label{eq:fh}
\end{equation}
One could consider a 
hamiltonian with a more general dependence on the $K_{ab}$, or on its
derivatives. However, at quadratic order, a term proportional 
to $K_{ab}K^{ab}$ is locally equivalent to $K^2$. This is because 
the Gauss-Codazzi equation (\ref{eq:gc})  
relates the difference $K^2 - K_{ab} K^{ab}$ 
to the surface scalar curvature ${\cal R}$, and
the hamiltonian 
constructed from ${\cal R}$ is a topological invariant, as follows from
the Gauss-Bonnet theorem. 

As always, the tangential variation of the hamiltonian is straightforward,
see Eq. (\ref{eq:tpda}).
We note that the normal 
variation of $H_{(n)}$, is given by
\begin{equation}
\label{eq:defk}
\delta_{\perp} H_{(n)} = \int_\Sigma dA \left\{ 
K^{n+1} \Phi  + n K^{n-1} \; \left(\delta_{\perp} 
K \right)   \right\} \,,
\end{equation}
where we have used Eq. (\ref{eq:sqk}) in the first
term. Using Eqs. (\ref{eq:defmet}) and (\ref{eq:defcur})
for the normal variation of the 
induced metric and the extrinsic curvature, respectively, we find
\begin{equation}
\delta_{\perp} K 
= -\Delta \Phi 
- K_{a b} K^{ab} \Phi \,.
\label{eq:dkab}
\end{equation}
Inserting this expression to Eq.(\ref{eq:defk}), and performing two integrations by parts
to collect the derivatives of $\Phi$ in a divergence, we obtain
\begin{eqnarray}
\delta_{\perp} H_{(n)} &=& \int_{\Sigma_0} dA
\left\{ -n\Delta K^{n-1} 
+ K^{n-1}(K^2 - n K_{a b} K^{ab})
\right\} \Phi
\nonumber \\
&-& n \int_{\Sigma_0} dA\,
\nabla_a \left\{  K^{n-1} \, \nabla^a\Phi - \nabla^a K^{n-1} \Phi 
\right\}\,.
\label{eq:varperpak}
\end{eqnarray}
We immediately identify the Euler-Lagrange derivative as
 \begin{equation}
{\cal E} (H_{(n)})
 = -n\Delta K^{n-1} 
+ K^{n-1}(K^2 - n K_{a b} K^{ab})\,.
\label{eq:elec}
\end{equation}
Generically, the Euler-Lagrange equation ${\cal E} =0$ is of second order
in derivatives of $K_{ab}$, so it is
 of fourth order in derivatives of 
the embedding functions ${\bf X}$.
In the exceptional case $n=1$, note that ${\cal E} = {\cal R}$ follows from
Eq.(\ref{eq:gc})  
which is of second 
order in derivatives of ${\bf X}$. 

The shape equation for the model described by 
Eq.(\ref{eq:helfrich}) takes the form
\begin{equation}
-2\alpha \Delta K + \alpha K (K^2 - 2 K_{ab} K^{ab} ) 
+ \beta {\cal R} + \mu K = P\,.
\end{equation} 

If the membrane possesses a
boundary, appropriate boundary conditions in the variational principle 
are identified by examining the divergence appearing in  Eq. (\ref{eq:varperpak}).
There are that $\Phi=0$, which kills both 
the second  term appearing in the divergence as well as the contribution 
from the derivative of $\Phi$ along ${\cal C}$ to the term 
proportional to $\nabla_a \Phi$. The normal derivative
$l^a\nabla_a\Phi$ remains. Thus we must set $l^a\nabla_a\Phi =0$ on ${\cal C}$.

We identify the operator $ S^a [\Phi]$   
introduced in Eq.(\ref{eq:perpdef}) which corresponds to $H_{(n)}$
as the `wronskian':
\begin{equation}
S^a_{(n)} [\Phi] = - n( K^{n-1} \; \nabla^a\Phi - \Phi \nabla^a
K^{n-1} )\,.
\end{equation}
In particular, if $\Phi$ corresponds to a background translation,
we have 
\begin{equation}
S^a{}_{(n)} = {\bf a}\cdot {\bf S}^a{}_{(n)} [{\bf n}] =
- n {\bf a}\cdot \left[ 
 K^{n-1} \, K^{ab} {\bf e}_b  - \nabla^a K^{n-1} {\bf n} 
\right]\,,
\end{equation}
where we exploit the Gauss-Weingarten equation
(\ref{eq:gw2})
to simplify the first term. We thus have from
Eq. (\ref{eq:pam}), the 
general expression for contribution to the stress tensor coming from $H_{(n)}$, 
\begin{equation}
{\bf f}^{a}_{(n)} = 
(n K^{n-1} \, K^{ab} - K^n \gamma^{ab})  {\bf e}_b  
- n \nabla^a K^{n-1} {\bf n} 
\,.
\label{eq:hnstress}
\end{equation}
Unlike the soap hamiltonian proportional to the area, 
the stress tensor
${\bf f}^{a}_{(n)}$ does possess a component normal to
the surface. If $n=1$, however, 
note that $f^a_{(1)}=0$ --- the corresponding stress is tangential.
If $\nabla_a K=0$ at any point, the normal stress vanishes there.
Note that the antisymmmetric part of $f^{ab}$  vanishes identically
for the geometrical hamiltonians we consider.

The hamiltonian $H_{(2)}$ is invariant under scale transformations. 
It is easy to see that this is equivalent to
\begin{equation}
 f^{ab}_{(2)} \gamma_{ab} = 0\,.
\label{eq:trss}
\end{equation}

For the complete Helfrich hamiltonian (\ref{eq:helfrich}), the stress
takes the form
\begin{equation}
\fl
{\bf f}^a = \left[
\alpha K (2 K^{ab} - K \gamma^{ab}) + \beta (K^{ab} - K
\gamma^{ab})
-\mu \gamma^{ab}\right] {\bf e}_b 
- 2 \alpha \nabla^a K {\bf n}\,.
\label{eq:hstress}
\end{equation} 
This is the principal result of this paper.
In general, the tangential stress $f^{ab}$ is neither homogeneous nor isotropic. Indeed,
it may vanish in places.
We note, in particular, that whereas the parameter $\mu$ 
is the thermodynamic tension, determining the response of the energy to a 
change in the membrane area, it is not the mechanical surface tension.

\section{Rotations and torque}
\label{rote}

Let us now consider an infinitesimal rotation, $\delta {\bf X}
= {\bf b} \times {\bf X}$. We have
$\Phi =  {\bf b}\cdot {\bf X}\times {\bf n}$ and 
$\Phi_a =  {\bf b}\cdot {\bf X}\times {\bf e}{}_a$,
and 
the variation (\ref{eq:defh}) of the hamiltonian associated with a
rotation
reduces to
\begin{equation}
\delta H_{\Sigma_0} = {\bf b}\cdot
\int_\Sigma dA\, \left[ {\cal E} (h)
\;  {\bf X}\times {\bf n}
-  \nabla_a {\bf m}^{a} \right]\,,
\label{eq:varang3}
\end{equation}
where the surface vector ${\bf m}^{a} $ is given by
\begin{equation}
{\bf m}^{a} = -
 {\bf S}^a[{\bf X}\times {\bf n}] - h\,{\bf X}\times  {\bf e}^{a} 
\,.
\end{equation}
The quantity ${\bf m}^a $ is identified as the torque 
about the origin acting on $\Sigma$.
Invariance of $H_{\Sigma_0}$ under rotations then implies, using 
${\cal E} = P$, 
\begin{equation}
\nabla_a {\bf m}^a = P \; {\bf X}\times {\bf n}\,.
\label{eq:mP}
\end{equation}
We isolate the contribution due to the
couple of ${\bf f}^a$ about the origin 

\begin{equation}
{\bf m}^{a} =
{\bf X} \times {\bf f}^a
+ {\bf s}^a \,,
\label{eq:angpi}
\end{equation} 
where 
 \begin{equation}
\label{eq:pi2}
{\bf s}^a{} 
= {\bf X}\times  {\bf S}^a [{\bf n}]  -
{\bf S}^a{}[{\bf X}\times {\bf n}] \,.
\end{equation} 
Neither ${\bf s}^a$ nor the couple due to ${\bf f}^a$ 
alone are conserved. 
An immediate consequence of Eqs. (\ref{eq:fE}), (\ref{eq:mP}), (\ref{eq:angpi}) is the relation
\begin{equation}
\nabla_a {\bf s}^a = {\bf f}^a \times {\bf e}_a\,.
\end{equation} 
We emphasize that this relation between ${\bf s}^a$ and ${\bf f}^a$
does not depend on the shape equation (\ref{eq:EP}); it holds 
for each term contributing to $H$.
Note that ${\bf s}^a$ does not involve derivatives 
in $K_{ab}$ other than those already contained in ${\bf f}^{a}$.

We can also expand ${\bf s}^a$ analoguosly to Eq.(\ref{eq:gep}) with tangential and normal 
projections, ${\bf s}^a = s^{ab} {\bf e}_b + s^a {\bf n}$. We have
 \begin{eqnarray}
{\nabla}_a s^{a} -  K_{ab}{}  s^{ab}  &=& \sqrt{\gamma} \; \epsilon_{ba}
\; f^{ab}\,, 
\label{eq:sem1} \\
\nabla_a  s^{ab} +  K^{b}{}_{a}   s^{a} &=& \sqrt{\gamma} \;
\epsilon_{ac} \; f^a \; \gamma^{cb} \,,
\label{eq:sem2}
\end{eqnarray} 
where we have used $\epsilon_{ab}= \epsilon_{\mu\nu\alpha} e^\mu_a e^\nu_b n^\alpha/\sqrt{\gamma}$,
and $\epsilon_{\mu\nu\alpha}$ is the Levi-Civita density.
The antisymmetric
part of $f^{ab}$, if present,  would contribute to Eq.(\ref{eq:sem1}).
For models that depend on polynomials of the extrinsic curvature, 
the symmetric part of $s^{ab}$ vanishes, 
and there is no torque about the normal, $s^a = 0$.

In particular, for the soap bubble the torque is simply that due to the 
couple of ${\bf f}^a$. The intrinsic torque associated with $H_{(n)}$, however, is
generally non-vanishing,
\begin{equation}
{\bf s}^{a}_{(n)} = 
 n K^{n-1}{\bf e}^a\times {\bf n} \,.
\label{eq:Mnab}
\end{equation}
Thus for  the Helfrich hamiltonian ${\bf s}^a$ is given by 
\begin{equation}
{\bf s}^{a} = 
 (2\alpha K  + \beta) {\bf e}^a\times {\bf n} \,.
\label{eq:Mab}
\end{equation}

\section{Global conservation}

Applying the divergence theorem to Eq.(\ref{eq:fE}) with ${\cal E} =P$ provides the global statement,
\begin{equation}
\label{eq:linang}
{\bf  F}({\cal C})  =  \int_{{\cal C}} ds\; 
 \; l_a \, {\bf f}^{a} = P \int _{\Sigma_0} dA\; {\bf n} 
\;.
\end{equation}
Recall that $l^a$ denotes the unit normal on ${\cal C}$
pointing out of $\Sigma_0$.
The total force exerted on the area element $\Sigma_0$ by the enclosed pressure $P$
is balanced by the internal forces exerted on the boundary curves ${\cal C}$.
If $P=0$, ${\bf F}({\cal C})$ vanishes. For spherical topology, or for a contractible 
loop ${\cal C}$ 
on a higher genus surface  this is also true for the individual closed ${\cal C}$. On  a non-contractible loop,
however, we have instead that ${\bf F}({\cal C})$ is a  non-vanishing constant vector. 
For a surface with the  topology of a torus, there are 
two distinct constant vectors corresponding to the 
two topologically distinct non-contractible circuits. 
Note that the modulus of this vector can be identified with the Casimir of the 
euclidean group associated with translations.
For a surface of genus $g$, there are $g+1$ distinct values.

The corresponding global statement
for the total torque acting on any closed loop ${\cal C}$ 
\begin{equation}{\bf M}({\cal C}) := 
\int_{\cal C} ds \;
l_a \, {\bf m}^{a} =
P
\int_\Sigma  dA 
\; {\bf X}\times {\bf n} 
\end{equation}
follows from Eq.(\ref{eq:mP}).

The total force ${\bf F} $ on a loop ${\cal C}$ on a soap film is given by
\begin{equation}
\label{eq:nglm}
{\bf F} ({\cal C}) =  - \mu \int_{{\cal C}} ds \; {\bf l}\,,
\end{equation}
where ${\bf l} = l^a {\bf e}{}_{a}$ is  the 
unit normal pointing out of  the surface 
at a given point on ${\cal C}$ treated as a space vector.
The corresponding torque acting on a loop ${\cal C}$ due to the enclosed membrane,
${\bf M} ({\cal C} )$ is given by
\begin{equation}
\label{ngam}
{\bf M} ({\cal C} ) =- {\mu} \int_{{\cal C}}  ds \,
{\bf X} \times {\bf l}\,.
\end{equation}

The  conservation law, Eq.(\ref{eq:fE}) has non-trivial consequences
which imply global constraints on the tangential projections $f^{ab}$.  
Let us dot Eq. (\ref{eq:fE}) with ${\bf X}$.
We have 
\begin{equation}
{\bf X} \cdot \nabla_a {\bf f}^a = 
\nabla_a ({\bf X}\cdot {\bf  f}^{a}) 
- {\bf e}_a \cdot {\bf f}^a =-  P \, {\bf X}\cdot {\bf n} 
\,.
\end{equation}
We note that ${\bf e}_a\cdot {\bf f}_b = f_{ab}$.
We integrate over $\Sigma$ to obtain the global constraint on the 
trace of $f^{ab}$,
\begin{equation}
\int_\Sigma dA
\;  f ^a{}_a =  - 3 P \; V \,.
\label{eq:intf}
\end{equation}
We have used  the representation
\begin{equation}
V = {1\over 3} \int_\Sigma dA \; {\bf X}\cdot {\bf n} 
\end{equation}
for the enclosed volume.
In addition to Eq. (\ref{eq:intf}),
there is the global constraint
\begin{equation}
\int dA \; K_{ab} \; f^{ab} =  P\; A\,,
\end{equation} 
obtained directly by integrating the projection (\ref{eq:em1})
over the surface.

For a soap bubble, the integrability condition Eq.(\ref{eq:intf}) gives
\begin{equation}
2 \mu A = 3 P V\,.
\label{eq:AV}
\end{equation}
This also follows as a consequence of stationarity of extremal configurations with respect to scaling. We have

\begin{equation}
\mu A [\lambda {\bf X}] -  P V[\lambda {\bf X}] = \lambda^2 \mu A[{\bf X}]- \lambda^3 \; P \;V[{\bf X}]\,,
\end{equation}
which is extremized, with $\lambda = 1$,  when Eq.(\ref{eq:AV}) is
satisfied.

For the Helfrich hamiltonian, the integrability condition Eq.(\ref{eq:intf}) gives
the well known scaling identity (see \cite{Sei:97} Eq.(3.10))
\begin{equation} \beta M + 2 \mu A = 3 P V\,.
\end{equation}
Note that the scale invariant $H_{(2)}$ does not contribute.

\section{Perturbations}

As an application of the conservation law 
(\ref{eq:fE}) let us examine briefly perturbations about equilibrium.
We will evaluate $\delta {\cal E}$ using our knowledge of the stress tensor and
compare our result with the one obtained using a `direct' approach\cite{Hel.OuY:89}. 
In particular the tangential 
perturbation is simply $\delta_\parallel {\cal E} (h) = \Phi^a \partial_a {\cal E} (h)$.
For the normal variation, it is  convenient to  consider 
the linearization of (\ref{eq:fE}): 
\begin{equation}
\delta_\perp \; \nabla_a \; {\bf  f}^{a} =  \; 
\left( \delta_\perp {\cal E} (h) \right)  \;  {\bf n} + 
{\cal E} (h)   \;  \delta_\perp \; {\bf n}\,.  
\label{eq:fEl}
\end{equation}
To express this equation in a more useful form  we can use
$\delta_\perp {\bf n} = -  ( \nabla_a \Phi ) {\bf e}^a $,
together with
\begin{equation}
\delta_\perp \; \nabla_a {\bf  f}^{a} =  \nabla_a  \delta_\perp {\bf  f}^{a}
- \left[ \nabla_a \left( K \Phi  \right) \right] {\bf f}^a\,,
\end{equation}
where we have used for the normal deformation of the 
connection, $ \delta_\perp \Gamma^b{}_{ab} = 
\nabla_a \left( K \Phi  \right) $. 
Now, taking projections of Eq. (\ref{eq:fEl}),
we obtain
the ``linearized Bianchi identity"
\begin{equation}
\left[ \nabla_a  \; \delta_\perp {\bf  f}^{a} \right] \cdot {\bf e}_b
-  f_b{}^a  \nabla_a \left( K \Phi  \right) 
= - ( \nabla_a \Phi ) {\cal E} (h)\,,
\end{equation}
and 
\begin{equation}
\left[ \nabla_a  \; \delta_\perp {\bf  f}^{a} \right] \cdot {\bf n}
- f^a \nabla_a \left( K \Phi \right)  = \delta_\perp {\cal E} (h)\,.
\end{equation}
The task of computing the linearization of the Euler-Lagrange
derivative is reduced by one order: one needs only to compute the
linearization of the stress tensor.

Let us introduce the quantities
\begin{equation} p_a = {\bf X}\cdot {\bf e}_a\,, \quad q ={\bf X}\cdot {\bf n}\,.
\end{equation}
Note that $p_a= \nabla_a {\bf X}^2/2$ is a gradient.
The Gauss-Weingarten equations can be cast in the alternative form
\begin{eqnarray}
\nabla_a p_b &=& - K_{ab}\, q + \gamma_{ab}\,,\\
\nabla_a q &=& K_{ab}\,  p^b\,.
\end{eqnarray}
Consider an infinitesimal dilatation under which $\delta {\bf X} = \lambda {\bf X}$. 
Then $q$ represents the normal component. As such we would expect
$\delta {\cal E} [ q] = inhom$, where $\delta {\cal E}$ is the 
linearized Euler-lagrange derivative and by `inhom' we mean terms which correspond to 
infinitesimal changes in the parameters of the model.
We note that under an infinitesimal rotation (three parameters),
$\epsilon \approx \epsilon^{ab} {\bf b} \times {\bf e}_a p_b $ satisfies    $\delta {\cal E}
=0$. Together with the (three parameter) 
infinitesimal translation $\epsilon = {\bf a}\cdot {\bf n}$
they exhaust the zero modes of $\delta {\cal E} = 0 $. A membrane with
translation symmetry along the $z-$ axis defines a unique loop on the 
orthogonal $x-y$ plane. The infinitesimal 
deformation of this loop induced by a rotation about this axis has $\epsilon \approx p$,
where $p= {\bf X}\cdot {\bf t}$ and ${\bf t}$ is the tangent vector to loop.
Thus, in this case, both projections of ${\bf X}$ satisfy the linearized 
equation. The consequences in the planar reduction of the Helfrich 
problem are discussed in detail in \cite{Arr.Cap.Chr.Guv:02}.

\section{Symmetries}

The approach we have developed is particularly appropriate when the 
membrane geometry possesses some degree of spatial symmetry. In particular, let us consider a membrane
that is axially symmetric. Then we can exploit
the integrated form  of the conservation law given by
Eq. (\ref{eq:linang}), and balancing forces on 
a circle of constant latitude we obtain a relation
between the appropriate components of the stress tensor
and the external pressure. When we specialize 
our considerations to the Helfrich hamiltonian (\ref{eq:helfrich}),
this relation reduces to the well known first
integral of the Helfrich shape equation for axially
symmetric configurations (see {\it e.g.} \cite{Zhe.Liu:93,Jul.Sei:94,
Boz.Sve.Zek:97}).

We choose the standard axially symmetric chart on $R^3$: 
$\rho, z, \varphi$. As coordinates on the surface we choose
the arc-length along the meridians, $l$, as well as 
$\varphi$.
The embedding of this geometry in $R^3$ can then be expressed as
\begin{equation}
\rho = R(l)\,, z = Z(l)\,.
\end{equation}
The line element induced on the surface is  
\begin{equation}
ds^2 = dl^2 + R(l)^2 d\varphi^2\,.
\end{equation}
The relation 
\begin{equation}
R'^2 + Z'^2 =1 
\end{equation}
defines $l$, where the prime denotes a derivative with respect to arc-length $l$.
A basis of tangent vectors for the surface 
adapted to this coordinate system is then given by
\begin{equation}
{\bf e}_ l = ( R', Z', 0)\quad 
{\bf e}_\varphi = (0,0,1)\,,
\end{equation}
and the outward pointing normal is
\begin{equation}
{\bf n} = (-Z', R',0)\,.
\end{equation}

Let us  consider a configuration with spherical topology.
We choose the loop ${\cal C}$ to coincide with a circle of fixed 
latitude, $z=$ constant. The vectors $l^a = (1,0)$, and 
$\epsilon^a=(0,R^{-1})$ denote the unit normal and tangent vectors respectively to this
circle on the surface. We note that 
 ${\bf l} = {\bf e}_a l^a = {\bf e}_l$.

We now examine the integrated form of the 
conservation law given by Eq.(\ref{eq:linang}).
We decompose the components of the
stress tensor, $f^{ab}$ and $f^a$, with respect to
the surface basis $\{ l^a , \epsilon^a \} $ as
\begin{eqnarray}
f^{ab} &=& f_{\perp\perp} l^a l^b + f_{||\,||}
(\gamma^{ab} - l^a l^b )\,, 
\label{eq:fab}\\
f^a & = & f_\perp l^a + f_{||} \epsilon^a\,,
\end{eqnarray}
where all four coefficients are independent of the angle $\varphi$. The potential off-diagonal term in 
$f^{ab}$ is zero by axial symmetry.
The term that appears on the l.h.s. of Eq. (\ref{eq:linang}) is
\begin{equation}
l_a {\bf f}^{a} = f_{\perp\perp} {\bf l} + f_\perp {\bf n}\,, 
\end{equation}
so that,
integrated around the loop, and using $ds = R(l) d\varphi$, we have
\begin{equation}
\int ds \, l_a {\bf f}^{a} = 
2\pi  R (
f_{\perp\perp}\,  \langle{\bf l} \rangle + f_\perp\, \langle {\bf n} \rangle)\,,
\end{equation}
where $\langle\cdot \rangle$ denotes an average over $\varphi$.
All but the $z$-components vanish. We have
\begin{equation}
\langle\,{\bf l}\, \rangle = (0, Z', 0)\,,\quad
\langle {\bf n}\rangle = (0,R',0)\,. 
\end{equation}
For the r.h.s. of Eq.(\ref{eq:linang}), this gives  
\begin{equation}
 2\pi P\int dl R \langle {\bf n} \rangle = 
\pi R^2 P \, (0, 1, 0)\,.
\label{eq:genm}
\end{equation}
Therefore Eq. (\ref{eq:linang}) reduces to  the relation
\begin{equation}
f_{\perp\perp} \;  Z' + f_\perp\; R' =
{ R \; P \over 2} \,.
\label{eq:nova}
\end{equation}
We emphasize that this expression does not 
depend on the details of the model, but only
on the assumption of axial symmetry and the
choice of the particular loop.

We define now the angle  $\Theta$ by $\sin \Theta = Z'$, $\cos\Theta= R'$.  Then
\begin{equation}
\tan \Theta = {d Z\over dR}
\end{equation}
When we decompose the extrinsic curvature as in 
Eq. (\ref{eq:fab}), we find
\begin{equation}
K_{\perp\perp} = \Theta'\,, \quad K_{||\,||}  = {\sin \Theta\over R}\,, 
\end{equation}
and for the mean extrinsic curvature,
\begin{equation}
K = \Theta' + {\sin \Theta \over R}\,.
\end{equation}

For a soap bubble, the relevant components of the 
stress tensor are, from Eq. (\ref{eq:lddng}), 
\begin{equation}
f_{\perp\perp} = - \mu\,, \quad f_\perp =0\,,
\label{eq:fsp}
\end{equation}
 so that the relation (\ref{eq:nova}) reduces to
\begin{equation}
\sin\Theta =   - R/ R_0\,, 
\end{equation} 
where $R_0 = 2\mu/P$.
The unique solution which is regular at the poles is a circle of radius $R_0$.

Let us now consider a membrane described by 
the Helfrich hamiltonian (\ref{eq:helfrich}).
As follows from Eq. (\ref{eq:hnstress}) for ${\bf f}^a{}_{(n)}$ we have 
\begin{eqnarray}
f_{(2)}^{\perp\perp} &=&  K ( 2 K_{\perp\perp} - K ) = 
(\Theta' + {\sin \Theta\over R}) 
( \Theta' - {\sin \Theta\over R} ) 
\,,
\nonumber \\
f_{(2)}^\perp &=& - 2 K' = - 2 \left (\Theta' +
{\sin \Theta \over R} \right){}'  \,,
\nonumber \\
f_{(1)}^{\perp\perp} &=& K_{\perp \perp} - K =  - \sin \Theta / R\,, \nonumber \\
f^\perp_{(1)} &=& 0\,.\nonumber
\end{eqnarray}
Therefore using these expressions, together with
the ones in (\ref{eq:fsp}), the relation (\ref{eq:nova})
gives
\begin{eqnarray}
-2 \alpha \cos\Theta \left(\Theta' + {\sin \Theta\over R}\right)' 
&+& \alpha (\Theta' + {\sin \Theta\over R}) 
(\Theta' - {\sin \Theta\over R}) \sin\Theta 
\nonumber \\
&-& \beta {\sin^2\Theta\over R}  - \mu \sin\Theta
=  {P R \over 2}  \,.
\label{eq:axeq}
\end{eqnarray} 
This coincides with the first integral for the axysymmetric 
shape equation obtained in \cite{Zhe.Liu:93}, and which is studied 
{\it e.g.} in \cite{Liu.Hai.Liu.OuY:99}. Our
approach should be
useful in providing an interpretation for an additional 
constant of integration $C$ that appears in these works.

Had we taken another loop, for example, the closed loop running along the meridan 
($\rho=0$) we find that we obtain a not so useful integral identity.

\section{Remarks}

In this paper, we have exploited  a combination of variational principles,
and conservation laws to examine the physics of 
two-dimensional surfaces which are described, at a mesoscopic level,
by an effective hamiltonian which depends locally on the 
surface geometry. In the case of the Helfrich hamiltonian, using
this approach, we have derived both the stress tensor and the
torque, associated with translational and rotational invariance,
respectively. The shape equation, which
determines equilibrium configurations, is now identified as one element 
of a conservation law.

The shift in focus onto the stresses in the membrane provides both
an intuitive description of equilibria and a useful calculational tool. 
We demonstrated this in the context of axially symmetric configurations. 
Using the integrated form of the conservation law for the stress tensor,
the first integral of the shape equation associated with 
this symmetry was obtained in an surprisingly economical way. 

We have explored various applications of this framework: 
a membrane with an exposed free boundary in \cite{Cap.Guv.San:02a};
and the adhesion of a membrane onto a substrate in \cite{Cap.Guv:02b}.
While both problems have been addressed previously in the axisymmetric context, the
boundary conditions associated with the geometry 
makes them very awkward to handle when this symmetry is relaxed.
We show how our framework is well suited not only to 
identifying these boundary conditions in this case, but also to 
providing a physical interpretation for them.
In a forthcoming paper, we will examine in detail how to approach
perturbation theory in the same framework.

\ack

We thank C. Chryssomalakos for his very useful comments.
RC is partially supported by CONACyT under grant 32187-E.
JG is partially supported by CONACyT 
under grant 32307-E and DGAPA at UNAM under grant IN119799.
We thank an anonymous referee for bringing Evans's early work to our attention.

\end{document}